\documentclass[12pt]{iopart}
\usepackage{graphicx}
\usepackage{iopams}  
\usepackage{hyperref}
\usepackage{xcolor}
\usepackage{tikz}
\usepackage{mathrsfs} 
\usetikzlibrary{decorations.markings,decorations.pathmorphing}
\usetikzlibrary{angles,quotes} 
\usetikzlibrary{arrows.meta} 

\newcommand{\calI}{\mathscr{I}} 
\tikzset{>=latex} 
\tikzstyle{singularity}=[line width=0.6,decorate,
                         decoration={zigzag,amplitude=2,segment length=6.17}]

\begin{document}

\title[The Synge G-Method:  cosmology, wormholes, firewalls, geometry]{The  Synge G-Method:  Cosmology, Wormholes, Firewalls, Geometry}

\author{G F R Ellis$^1$ and D Garfinkle$^2$ }


\address{$^1$ Mathematics and Applied Mathematics Department, University of Cape Town, Cape Town 7700, South Africa.

email: george.ellis@uct.ac.za}

\address{$^2$ Dept. of Physics, Oakland University, Rochester, MI 48309, USA

email: garfinkl@oakland.edu}

\vspace{10pt}
\begin{indented}
\item[]November 2023
\end{indented}

\begin{abstract}
 Unphysical equations of state result from the unrestricted use of the Synge G-trick of running the Einstein field equations backwards;  in particular often this results in $\rho + p < 0$ which implies negative inertial mass density, which does not occur in reality. This is the basis of some unphysical spacetime models including phantom energy in cosmology and traversable wormholes. 
 
 The slogan ``ER = EPR'' appears to have no basis in physics and is merely the result of vague and unbridled speculation.  Wormholes (the ``ER'' of the slogan) are a mathematical curiosity of general relativity that have little to no application to a description of our universe.  In contrast quantum correlations (the ``EPR'' of the slogan) are a fundamental property of quantum mechanics that follows from the principle of superposition and is true regardless of the properties of gravity.

 The speculative line of thought that led to ``ER = EPR'' is part of a current vogue for anti-geometrical thinking that runs counter to (and threatens to erase) the great geometrical insights of the global structure program of general relativity.
 
\end{abstract}

\section{Introduction}

J L Synge many years ago showed how a simple process (his ``G-Method'') could lead purely by differentiation to exact solutions of the Einstein Field Equations (``EFE''). However this often leads to a negative inertial mass density, hence they are unphysical solutions of the EFE.  In this paper we discuss how this occurs both in cosmology, and in proposals made by members of the string theory community for the existence of ``traversable wormholes''.   We argue they can neither come into existence by physically plausible processes, nor - even disregarding that problem - can they actually exist .

There is a tension as regards gravitational theory between those who come from the general relativity side, who by and large approach the subject in a geometric fashion \cite{Hawking and Ellis (2023)}, and those who come from the particle physics/string theory side, who by and large treat  it non-geometrically \cite{Weinberg (1972)} except sometimes for the use of some causal diamonds to represent causal relations. A key feature is that for the former, the ``force of gravity'' is a fictitious force resulting from the effects of the curvature of spacetime seen in an accelerated frame, so that while the idea of quantising the gravitational force may be a useful effective theory, at a fundamental level it does not make sense. For the latter, gravity is just another force, to be quantised like all the others. This is a key aspect of the difference between geometric and non-geometric approaches.

This paper discusses \S2, The Synge Trick and Energy Conditions, with applications to cosmology; \S3, Wormholes in General Relativity; \S4, AdS/CFT, Wormholes and EPR; \S5, Horizons, Firewalls and Atlases; \S6, Geometric and non-geometric approaches. 

\section{The Synge Trick and Energy Conditions}
There are two quite different ways of solving EFE: the standard way, and the Synge G-Method (\S 2.1).  The latter is very easy to do but often results in unphysical matter tensors, and violates stability conditions  (\S 2.2). However it can be useful in inflationary contexts (\S2.3), but can lead to problems with dark energy phenomenology (\S2.4).

\subsection{Two ways of solving the EFE}
Many years ago, J. L. Synge  \cite{Synge (1961)} pointed out an inverse way of solving the  the EFE
\begin{equation}\label{eq:EFE}
    R_{ab}- \frac{1}{2}R g_{ab} + \Lambda g_{ab} = \kappa T_{ab}, \,\,\Lambda_{,c}=0,
\end{equation}
 where $\kappa > 0$ is the gravitational constant, with integrability condition
\begin{equation}\label{eq:conservation}
      (R^{ab}- \frac{1}{2}R g^{ab})_{;b} = 0 \Leftrightarrow  (T^{ab})_{;b} = 0.
 \end{equation}
These are usually solved by the following process, from right to left: 
\begin{itemize}
    \item Assume a specific form for the energy-momentum tensor $T_{ab}$ (a vacuum, perfect fluid, electromagnetic field, scalar field, and so on), \item  Impose some symmetries on $g_{ab}(x^i)$ to represent an interesting physical context (a cosmological model, a black hole, a gravitational wave, and so on),  possibly linearising about a space with exact symmetries, \item  Calculate the Ricci tensor $R_{ab}(x^i)$ and Ricci scalar $R(x^i)$ as a function of the metric $g_{ab}(x^i)$, its inverse $g^{cd}(x^i)$, and their first and second derivatives,  \item   Substitute in (\ref{eq:EFE}) to obtain partial differential equations (PDE) for the metric and any matter fields.  Also impose any additional PDE that the matter fields themselves have to satisfy (e.g. Maxwell's equations for the electromagnetic field).
    
    \item Solve those PDE for $g_{ab}(x^i)$ and any associated matter fields. 
\end{itemize}
However one can proceed in the opposite manner, from left to right:
\begin{itemize}
    \item Assume a specific spacetime geometry $g_{ab}(x^i)$,
    \item Calculate the Ricci tensor $R_{ab}(x^i)$ and Ricci scalar $R(x^i)$ from $g_{ab}(x^i)$, 
    \item Substitute in the EFE (\ref{eq:EFE}) to determine $T_{ab}(x^i)$.
    \item The solution $\{g_{ab}(x^i),T_{ab}(x^i)\}$ will automatically satisfy (\ref{eq:conservation}).
\end{itemize}
This is the Synge G-method, a  simple process that generates an exact solution of the EFE without the need to solve any partial differential equations.

Note that to apply the Synge G-method without any restrictions is to deprive the Einstein field equations of all content.  To see why, consider the following analogy with Newton's second law ${\vec F} = m {\vec a}$.  Pick any particle trajectory ${\vec x}(t)$ and calculate ${\ddot {\vec x}}(t)$.  Now assert that you have a solution to ${\vec F} = m {\vec a}$ where the force is given by $m{\ddot {\vec x}}(t) $.  Since this can be done with any trajectory, the statement ``I have a trajectory that is a solution of ${\vec F} = m {\vec a}\;$'' then has no more content than the statement ``I have a trajectory.''  Similarly, if one uses the Synge G-method without restriction, the statement ``let $(M,{g_{ab}})$ be a spacetime satisfying the Einstein field equations'' has no more content than the statement ``let $(M,{g_{ab}})$ be a spacetime.''

However, the Synge G-trick does become meaningful (and indeed quite useful) with the imposition of only a causality condition (like the presence of a Cauchy surface or the absence of closed timelike curves) and an energy condition (to be discussed below). Indeed it is the great triumph of the global structure program (see \cite{Hawking and Ellis (2023)} and references therein) that powerful results can be obtained using only such mild restrictions.

The main problem with the unrestricted G-method is that the stress-energy tensor $T_{ab}(x^i)$ determined this way will in general not correspond to any known form of matter, thus it will be unphysical. Furthermore it will in general not satisfy the stability conditions that are required for a solution to be physically meaningful \cite{Bishop and Ellis (2020)}.

\subsection{Energy conditions and stability conditions}
In many cases a `perfect fluid' description is used in general relativity, where the stress-energy tensor takes the form
\begin{equation}\label{eq:PF}
    T_{ab} = (\rho + p) u_{a}u_{b} + p g_{ab},\,\,u^a = dx^a/d\tau,\,\,u^a u_a = -1 
\end{equation}
where $\rho(x^i)$ is the matter density, $p(x^i)$ its pressure in units where the speed of light $c=1$, and $\tau$ is proper time along the fluid world line $x^a(\tau)$.  

In the case of a perfect fluid (\ref{eq:PF}), the conservation equations (\ref{eq:conservation}) become the energy equation (\ref{eq:energy}) and momentum equation (\ref{eq:momentum}):
\begin{eqnarray}
    \dot{\rho} + (\rho +p) \Theta = 0, \label{eq:energy}\\
    (\rho + p) \dot{u}^a +\nabla p =0 ,\label{eq:momentum}
\end{eqnarray}
where $\Theta: = u^a_{\,\,\,;a}$ is the fluid expansion, $\dot{u}^a := u^a_{\,\,;b}u^b$ its relativistic acceleration, and $\nabla p :=(g^a_{\,\,b} + u^a u_b)p_{;b}$ the spatial gradient of the pressure  \cite{Ehlers (1993),Ellis (1971)}. 

The \textit{weak energy condition} is that 
\begin{equation}\label{eq:Weak}
    \rho \geq 0,\,\,(\rho+p)\geq 0.
\end{equation}
No known matter violates  (\ref{eq:Weak}). If it were  violated, the inertial mass density would be negative: \begin{equation}\label{eq:NWE}
    \rho_{inert} := \rho +p <0 \Rightarrow p < 0
\end{equation}
which leads to unphysical behaviour of two kinds.
\begin{itemize}
    
\item From (\ref{eq:momentum}), if you push the fluid to the left, it goes to the right, unlike any ordinary matter. 
If $p = w \rho$ where  $w$ is  constant or almost constant, the speed of sound $c_s$ is given by $ c_s^2 =\frac{dp}{d\rho} = w 
$ \cite{Ellis et al (2007)}.     Then (\ref{eq:Weak}) can be rewritten as
    \begin{equation}
        (1+w) \rho  <0 \Rightarrow 1 + c_s^2 < 0.
    \end{equation}
    Thus the speed of sound is imaginary and mechanical instability arises (perturbation gives  exponential expansion or collapse rather than oscillations). Also  runaway motion occurs \cite{Bonnor (1989)}.    Thomas Gold suggested such runaway linear motion could be used in a perpetual motion machine if converted to circular motion by attaching positive and negative masses to the rim of a wheel (\cite{DeWitt-Morette and Rickles (2017)}:159), hence it cannot occur. 

 \parindent 20.0pt \indent This is to be distinguished from gravitational instability, which occurs whenever the active gravitational mass  is positive: $\rho_{grav} := (\rho + 3p) > 0$. This is obeyed by all baryonic matter and plays a key role in structure formation in the universe, but can be violated by effective scalar fields, as in the inflationary era in the early universe \cite{Peter and Uzan (2009), Ellis and Uzan (2017)} .
    \item     From (\ref{eq:energy}), if the fluid expands and $\rho_{inert}<0$, the density goes up, not down, contrary to all experimental evidence for known matter, and the pressure has to become even more negative to preserve inequality (\ref{eq:NWE}). This unphysical behaviour results in the prediction of Sudden Singularities \cite{Barrow (2004)}: the fluid density diverges in a finite time as it expands. Nothing we know behaves like this.  If it were true, you could in principle obtain an infinite amount of energy from such a fluid in a finite time - which is unphysical \cite{Ellis et al (2018)}. 
\end{itemize}
 In summary: the standard method produces solutions of the EFE that correspond to the kind of matter we know exists. Use of the Synge G-method results in ``solutions'' that in general do not correspond to any kind of matter we have ever encountered, and arguably  allow perpetual motion - that is, they represent unphysical behaviour.

\subsection{Inflationary potentials and dark energy phenomenology}
An interesting variant of the G-method is that when inflation takes places in cosmology \cite{Peter and Uzan (2009), Ellis and Uzan (2017)}, driven by a scalar field $\phi$ with potential $V(\phi)$, one can choose an (almost) arbitrary scale factor $a(t)$ and then run the EFE backwards to determine a potential $V(\phi)$ that will lead to that scale factor  $a(t)$
 \cite{Ellis and Madsen (1991)}. Indeed this is what is done in practice in many inflationary cosmology studies \cite{Martin et al (2014)}. Such potentials will in general not be directly related to any deeper physical models: they are simply designed to give the desired dynamic result. They are simply a phenomenological fit of the ``matter'' potential $V(\phi)$ as required to give the desired inflationary dynamics $a(t)$.

\subsection{Dark energy equations of state}
Famously, the late time universe has been determined to be accelerating, where the  scale factor $a(t)$ obeys  $\ddot{a}(t) >0$ \cite{Peter and Uzan (2009),Ellis and Uzan (2017)}. This might be due to a cosmological constant but many possibilities of ``dark energy'' are also being explored, with a phenomenogical equation of state $w(t)$ defined by 
\begin{equation}
    p(t) = w(t) \rho(t).
\end{equation}
Here $w(t)$ is to be determined from the observational relation $r_O(z(t))$ between observed  area distance $r_O(t)$ and redshift $z(t)$  where $t$ is the time of emission  of light \cite{Ellis (1971)}. This relation determines $a(t)$ and hence,  on using the EFE, $p(t)$ and so $w(t)$, and many investigations are under way to do this. These are examples of the Synge G-method: the logic is, 
$r_O(z(t)) \Rightarrow a(t) \Rightarrow w(t). $ Observational analyses usually
include cases where $w < -1$ (called phantom energy), that is $\rho_{inert} < 0$ \cite{Melchiorri et al (2003)},\cite{Vazquez et al (2012)}). This runs into the same kind of stability issues as just discussed \cite{Carroll et al (2003), Vikman (2004)}. In our view this domain of the parameter space for $a(t)$ should be rejected as being unphysical.

\section{Wormholes in General Relativity}

Before delving into the exotic, implausible claims that have been recently made about wormholes, it is helpful to go over what general relativity has to say about wormholes.

To begin, note that general relativity has an initial value formulation: that is we pick the initial configuration and the theory tells us how it evolves.  But because general relativity is a geometric theory it has a degree of freedom not present in our other theories: the topology of space.  
In all our other theories the topology of space is $R^3$.  But in general relativity other topologies are allowed.  In particular, the topology of space in the extended Schwarzschild spacetime is ${S^2} X R$, that is there are two asymptotically flat spaces connected along a sphere. This is what is referred to as a ``wormhole''.

Though we have referred to spatial topology as a ``degree of freedom'' it is so only in a very restricted sense: the topology of space remains whatever it was at the initial time, as shown in the Geroch no topology change theorem:  In classical GR, no topology change can occur in the development of initial data on a spatial hypersurface  without causal violations or a singularity occurring \cite{Geroch (1967)}. An example is that the spatial topology of standard cosmological models cannot change \cite{Ellis (1987)}.

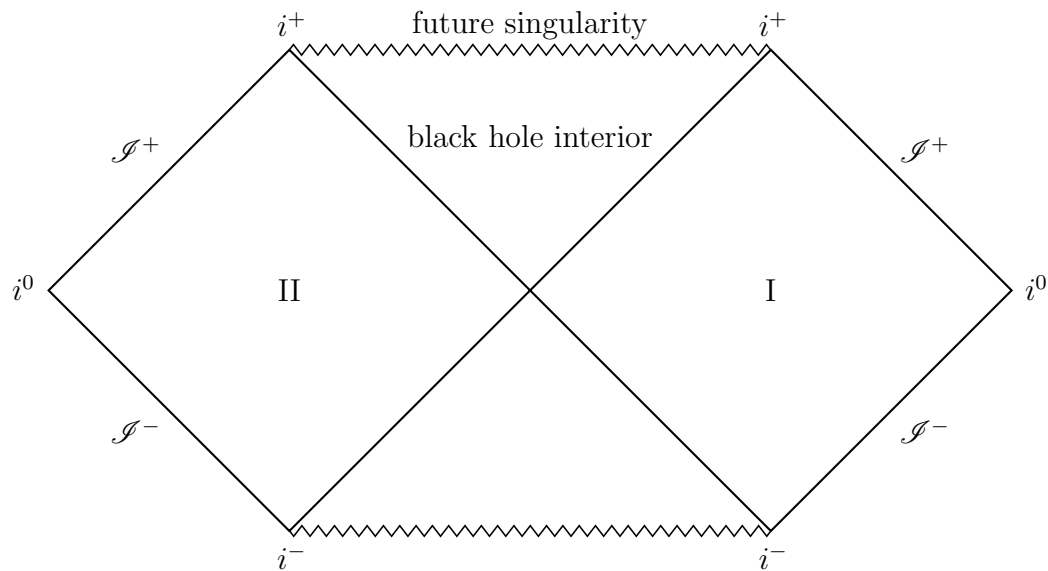
\begin{figure}
\centering
\begin{tikzpicture}[scale=3.2]
  
  \coordinate (-O) at (-1, 0); 
  \coordinate (-S) at (-1,-1); 
  \coordinate (-N) at (-1, 1); 
  \coordinate (-W) at (-2, 0); 
  \coordinate (-E) at ( 0, 0); 
  \coordinate (O)  at ( 1, 0); 
  \coordinate (S)  at ( 1,-1); 
  \coordinate (N)  at ( 1, 1); 
  \coordinate (E)  at ( 2, 0); 
  \coordinate (W)  at ( 0, 0); 
  \coordinate (B)  at ( 0,-1); 
  \coordinate (T)  at ( 0, 1); 

  \draw[singularity] (-N) -- node[above] {future singularity } (N);
  \draw[singularity] (S) --   (-S);
  \draw[thick] (-N) -- (-E) -- (-S) -- (-W) -- cycle;
  \draw[thick] (N) -- (E) -- (S) -- (W) -- cycle;
  
  \node[inner sep=2] at (-O) {II};
  \node[inner sep=2] at (O) {I};
  \node[inner sep=2] at (0,0.64) {black hole interior};

  \node[above=1,left=1] at (-2,0) {$i^0$};
  \node[above=1,right=1] at (2,0) {$i^0$};
  \node[right=1,below=1] at (-S) {$i^-$};
  \node[right=1,above=1] at (-N) {$i^+$};
  \node[right=1,below=1] at (S) {$i^-$};
  \node[right=1,above=1] at (N) {$i^+$};
  \node[below left=-1] at (-1.5,-0.5) {$\calI^-$};
  \node[above left=-1] at (-1.5,0.5) {$\calI^+$};
  \node[above right=-1] at (1.5,0.5) {$\calI^+$};
  \node[below right=-1] at (1.5,-0.5) {$\calI^-$};

\end{tikzpicture}
\caption{Penrose diagram of the maximally extended Schwarzschild spacetime.  The two asymptotically flat regions, I and II, share a common black hole interior}
\label{ExtendedSchwarzschild}
\end{figure}

Thus we either live in a space of exotic topology or we don't.  In either case we have no power to change this.  Nonetheless, one might think that if we did happen to live in a space with exotic topology, we might explore this space at our leisure and map out its topology.  That we can't do this is shown in two sets of results: Gannon's theorem and the topological censorship theorems.  Gannon's theorem \cite{Gannon (1975)} says that the evolution of initial data with exotic topology results in the production of a spacetime singularity.  Thus, for example the initial data for the extended Schwarzschild spacetime evolves to the extended Schwarzschild spacetime: a black hole with a spacetime singularity in its interior. 

The topological censorship theorems \cite{Friedman et al (1993)} say that no observer can visit more than one part of the exotic topology and thus that no observer will notice that they live in a space with exotic topology.  Thus, for example in the extended Schwarzschild spacetime (Figure \ref{ExtendedSchwarzschild}) observers either remain in asymptotically flat region I or remain in asymptotically flat region II, or fall into the black hole.  Observers who remain in asymptotically flat region I have no contact with those in asymptotically flat region II and have no way of knowing that they are there.  Or to put it another way, these observers have no way of knowing that the black hole in their space is really a wormhole.  The only observable consequence of the exotic topology is that observers from the two regions can each jump into the black hole and briefly meet each other on their way to being crushed by the spacetime singularity (and without any way to tell the people back home what happened).   

\section{AdS/CFT, Wormholes and EPR}
Following a paper by Maldacena and Susskind \cite{Maldacena and Susskind (2013)}, a huge literature has developed around the slogan ``ER = EPR'' linking ``Einstein-Rosen Bridges"  (ER: the idea of traversable wormholes \cite{Einstein and Rosen (1935)}) on the one hand, and the idea of quantum entanglement following on the groundbreaking paper by Einstein, Podolsky, and Rosen  (EPR: \cite{Einstein et al (1935)}) on the other (the idea of quantum entanglement was also proposed by Schr\"{o}dinger \cite{Schrodinger (1935)}). It is often alleged that experimental demonstration of the existence of entanglement, as has been done,  proves spacetime wormholes exist. It is even alleged that such wormholes have been created in a quantum computer, whose operation is based in entanglement. 

However traversable wormholes do not occur naturally in GR: the wormhole in the maximal Schwarzschild extension (\cite{Hawking and Ellis (2023)}:154, see Fig.1) is non-traversable, as is also the case in the  the maximal Reissner-Nordstr\"{o}m (\cite{Hawking and Ellis (2023)}:158) and Kerr solution (\cite{Hawking and Ellis (2023)}:165). The Weak Energy Condition is violated in the case of traversable wormholes \cite{Friedman et al (1993)} - a result known as ``topological censorship''. Sure one can construct a traversable wormhole by choosing a suitable metric and using the Synge G-Method. The resulting spacetime will be non-physical, as it will violate the weak energy condition (\ref{eq:Weak}).

Here we discuss, the claims made (\S 4.1), Can they come into existence? (\S 4.2), Can they exist? (\S 4.3), A simulation is not the thing (\S 4.4), and Is ER = EPR true? (\S 4.5).

\subsection{The claims made}

The paper \cite{Maldacena and Susskind (2013)} (hereafter MS) considers two different wormhole spacetimes: (1) the Schwarzschild-anti-de Sitter spacetime, and (2) the spacetime for black hole pair creation by a magnetic field treated in \cite{Garfinkle and Strominger (1991)}.  

In both cases these are non-traversable wormholes.  That is, just as in the extended Schwarzschild spacetime, these are two spaces where their black holes share a common interior.  Thus when MS consider signals sent from an observer in one space to an observer in the other space, these messages can only be received when the second observer jumps into the black hole.

In both cases there are quantum correlations between the two sides of the wormhole spacetime, but in each case these correlations seem to have nothing to do with the fact that the spacetime contains a wormhole: In Schwarzschild-anti-de Sitter spacetime the wormhole is eternal and the quantum correlations are there simply because the vacuum state has quantum correlations.  Or to put it another way, there are quantum correlations between the two halves of the wormhole for the same reason that there are quantum correlations between the two Rindler wedges in Minkowski spacetime.  In the black hole pair creation case, the correlations are there because pair creation produces correlations: again nothing to do with the fact that there is a wormhole.

Thus, in terms of their slogan ``ER=EPR" up to that point MS had produced two examples of ER that also happened to be EPR, though not for any reason to do with the fact that they were ER.  However, MS then went on to make (what even they considered at the time to be) the bold speculation that all cases of EPR (i.e. quantum correlations) are dual (in the sense of AdS/CFT) to ER (i.e. wormhole) spactimes.  We will attempt to evaluate this claim in subsequent sections, but for now we note (as a tidbit of sociology of science) that in the years since 2013, ER=EPR seems to have gone from bold speculation to received wisdom without any solid arguments on its behalf having ever been offered.

In contrast to MS, \cite{Maldacena et al (2023)} consider traversable wormholes.  The matter needed to hold these wormholes open necessarily violates the weak energy condition.
Similarly the wormholes in \cite{Gao et al (2017)} are also traversable and also involve matter that violates the weak energy condition.

\subsection{Can they come into existence?} Do wormholes satisfy the conditions of Assembly Theory \cite{Sharma et al (2023)}?  That is, can they come into existence by an assembly process from pre-existing components ?

Classically: no. In general relativity theory, not only  can't wormholes be assembled from existing components, but also they can't be made by any process at all: either 
a wormhole has existed forever or it won't exist at all. 
 Certainly a wormhole exists in the maximally extended Schwarzschild solution (\cite{Hawking and Ellis (2023)}:154), but that wormhole has existed for ever, and so does not relate to any process of astrophysical formation as understood either from the general relativity side \cite{Penrose (1965)} or the astrophysical side \cite{Begelman and Rees (2020)}. In realistic astrophysical processes, the potential wormhole that exists in the maximal Schwarschild solution as represented in the Kruskal diagram is cut off by the infalling matter that creates the black hole (\cite{Hawking and Ellis (2023)}:300,302,309; \cite{Hawking (1975)}:Figs 2,3). As for the original Einstein-Rosen paper \cite{Einstein and Rosen (1935)}, it involves modifying the EFE so it is not in fact a General Relativity solution.

Quantum mechanically: maybe. Not through assembly but through quantum tunneling. The most reliable way to do calculations in quantum field theory is within a perturbative regime using Feynman diagrams.  But quantum wormhole production is a non-perturbative quantum tunneling process calculated using instantons (as e.g. in \cite{Garfinkle and Strominger (1991)}).  Here it is helpful to recall the limitations of such calculations: (1) since we don't have a quantum theory of gravity we don't know whether such a theory has a topology superselection rule (a quantum analog of the Geroch no topology change theorem) that would forbid the production of wormholes.  (2) we don't know for sure that the action of a Euclidean spacetime is the correct quantity to put in the quantum tunneling formula.  We do this because it is analogous to what does work in (non-gravity) quantum theories where we know how to do calculations.  (3) instanton calculations are semiclassical but also exponentially suppressed.  This means that the wormholes produced need to be sufficiently larger than the Planck mass that we can trust the calculation, but also sufficiently small that their production rate is not negligibly tiny.  That amounts to a wormhole mass of order say 10 Planck masses.  Note that such a mass is much too small for any of the thought experiments envisioned in MS.  So to perform any of their thought experiments they would need to first grow the wormhole to a much larger size.  However, since their thought experiments involve observers sending messages ``sufficiently early'' it is not clear which (if any) of those thought experiments are compatible with the wormhole production and growth process.

\subsection{Can they exist?}
So can they in fact exist? Insofar as traversable wormholes are solutions of the EFE, they are examples of use of the Synge G-method without restriction (\S2.1). Vacuum wormhole EFE solutions exist  (the Kruskal maximal extension of the vacuum spherically symmetric Schwarzschild solution is a wormhole) but are not traversable. Traversable wormholes are derived by assuming a traversable spherically symmetric geometry and then running the EFE backwards to determine whatever form of ``matter" allows such solutions to exist. This does not imply that such ``matter'' is physically plausible, or can indeed exist in reality. \cite{Morris and Thorne (1988)} acknowledge that traversable wormholes require unphysical matter, but press on anyhow suggesting they are a good tool for teaching general relativity.  We strongly disagree.  Maldacena \textit{et al} present a traversable wormhole in four dimensions in \cite{Maldacena et al (2023)}, but it is based in the concept of charged massless fermions - which do not exist in physical reality and would be subject to the Coleman-Weinberg instability. 
We conjecture that if they were indeed to exist despite these considerations, they would be highly unstable and subject to singularities such as ``sudden future singularities'' \cite{Barrow (2004)} in a finite time because of the nature of the ``matter'' involved.

\subsection{A simulation is not the thing}
 Recently some excitement was generated by a paper 
''Traversable wormhole dynamics on a quantum processor'' \cite{Jafferis et al (2022)}, claiming that traversable wormholes had been created by a quantum computer. But this was actually a simulation of (a system speculated to be the dual of) such wormholes, not the thing itself, and the press release proclaiming this event generated much incredulity\footnote{See \href{https://www.math.columbia.edu/~woit/wordpress/?p=13181}{This Week’s Hype} by Peter Wojt}. In any case a quantum computer generates a negligible gravitational field - not remotely large enough to create a black hole let alone any associated wormhole.

For what it's worth, we also note that computer simulations of classical non-traversable wormholes are done routinely and have been done for decades as a byproduct of a particular method for binary black hole simulations.\cite{Campanelli et al (2005)}  This is because this method (called the moving puncture method) chooses initial data where each of the black holes is connected to another space with initial data close to that of the extended Schwarzschild spacetime.  For the region outside the black hole horizon, these simulations with wormholes are in complete agreement with other binary black hole simulations that don't use wormhole initial data,\cite{Pretorius (2005)} thus once again showing that in GR you can't tell if there is a wormhole.

\subsection{Is ER = EPR true?}
The original EPR paper \cite{Einstein et al (1935)} as well as Schr\"{o}dinger's basic paper on entanglement \cite{Schrodinger (1935)} are  calculations in flat spacetime, as are subsequent discussions of quantum entanglement \cite{Horodecki et al (2009), Susskind and Friedman  (2014)}. They have nothing whatsoever to do with wormholes, as denoted by ER.  Experimental proofs that entanglement is possible over macroscopic scales, such as those that won the Nobel Prize in 2022, were also based on flat spacetime calculations. By contrast a standard wormhole solution of the  EFE represents the maximal spacetime curvature that is possible - spacetime is curved so highly by   matter   that both closed trapped surfaces and event horizons exist \cite{Penrose (1965),Hawking and Ellis (2023)}. These are  manifestly completely different spacetime situations both geometrically and physically. Entanglement (EPR) is possible in both cases: it does not require a spacetime wormhole (ER). 

Thus despite what has been claimed vociferously by some, ER = EPR is not true in general, and in particular experimental verification of quantum entanglement either in a laboratory or at macroscopic scales does not imply that spacetime wormholes exist.\footnote{One of us (GE) has been vociferously attacked on social media for denying that experimental proof of entanglement implies existence of spacetime wormholes.} How a wormhole is defined in a quantum gravity context, where a spacetime metric may not be defined, is not even clear.

\section{Horizons, Firewalls, and Atlases}

Where did the the ER = EPR conjecture originate? To fully understand this we have to consider the phenomenon of black hole evaporation. 
We discuss here in turn, Atlases and local wavefunctions (\S 5.1), Is there a Black Hole Information Paradox? (\S 5.2), 
Black hole evaporation (\S 5.3),
and The Firewall and ER = EPR (\S 5.4).

\subsection{Prolog: Atlases and local wavefunctions}
Preliminary to the subsequent discussion, we consider 
coordinate atlases and local wavefunctions.

\subsubsection{Atlases}
A crucial feature enabling global studies in GR was the introduction of coordinate atlases $\{ {\cal U}_\alpha,x_\alpha^i\}$ (\cite{Hawking and Ellis (2023)}:\S 2.1). The coordinate system $\{x_\alpha^i\}$ is valid only in the open set ${\cal U}_\alpha$, and the relevant manifold ${\cal M}$ is covered by a union $\cup_\alpha \{{\cal U}_\alpha \}$ of such coordinate neighbourhoods. Different coordinate systems for the same manifold are related by diffeomorphisms. Even as simple a space as the 2-sphere ${\cal S}^2$ cannot be covered by a single coordinate system.

\subsubsection{Local Wave functions}
A key puzzle for quantum theory is how a strictly linear theory - a wave function $|\psi (t) \rangle$ obeying the time-dependent first order Schr\"{o}dinger equation \cite{Laughlin and Pines  (2000)}
\begin{equation}\label{eq:Dirac}
    i \hbar \frac{\partial |\psi(t) \rangle}{\partial t} = \hat{H} |\psi (t) \rangle
\end{equation}
can underlie all the non-linear emergent dynamics we see around us all the time. It can be proposed that the same kind of analysis as in the GR case is needed: in complex systems, only local wave functions  $|\psi (t) \rangle_\alpha$ exist, each valid in a local region $\{{\cal U}_\alpha \}$ small enough that the dynamics is still linear, enabling physics to be locally unitary everywhere but linked in non-linear ways to create complex outcomes \cite{Ellis (2023)}. The implication is that in cases such as quantum cosmology or the emission of Hawking radiation,  one is not entitled to simply assume \textit{a priori} that a global wave function exists for the system concerned: this has to be justified. In the case of black holes forming by astrophysical processes, this is almost certainly not the case. The ``black hole information paradox'' assumes (1) unitary dynamics (\ref{eq:Dirac}),  perhaps expressed in terms of the associated Klein-Gordon equation, which in turn assumes (2) existence of a single relevant wave function $|\psi (t) \rangle$ in \textit{all} relevant domains. We will question both assumptions below.

\subsection{Is there a Black Hole Information Paradox?}

Black body radiation is emitted by black holes \cite{Hawking (1975)} irrespective of the matter out of which the black hole is formed. By the black hole uniqueness theorems, the result will be a Schwarzschild or Kerr black hole with closed trapped surfaces and with the interior hidden behind an event horizon (in astrophysical reality, charged black holes will not occur), The solution is fully characterised by just two parameters; the  mass $m$ and rotation parameter $a$. Thus all information about the matter that created the black hole, or anything that later falls into it, is completely lost to those who remain outside the black hole. This contradicts the intuition that different initial states should always lead to different final states, no matter how small the difference is. More formally, the assumption is made that the associated quantum evolution is unitary. But this is not true, either classically or quantum mechanically 

\subsubsection{Classically}

In the classical case, of course no black hole radiation takes place. Information is not conserved: if you drop anything in, there is a singular boundary to spacetime
and both matter and information get lost as they leave spacetime. No conservation law applies within spacetime because of this local boundary.

\subsubsection{Quantum Mechanically} In the quantum case, Hawking radiation is emitted \cite{Hawking (1975)} and information is lost in the sense that different starting states lead to the same outcome   \cite{Hawking (1976),Mathur (2009)}. The ``information paradox'' arises from the assumption that the dynamics is unitary.  One possible resolution of this paradox is that black hole evaporation can be regarded as a wave function collapse situation, not a unitary one, because the initial quantum state leads to a classical detection event - and any such quantum to classical transition is non-unitary  and information is lost \cite{Ellis (2012)}.  It is basically a measurement event (\cite{Aaronson (2016)}:\S1.2.1 2.). Furthermore while the calculation done uses different bases for the inward and outward wave functions, it nevertheless assumes a single wave function exists for the entire spacetime  - which may not be justified. There may be no such single wavefunction, just as there is no ``Wave Function of the Universe'' \cite{Ellis (2023)}. In particular this lack of a single wavefunction may be expected to be the true in the case of astrophysical black holes \cite{Begelman and Rees (2020)}, where manifestly the dynamics is very non-linear.  The underlying issue is whether these discussions are related to highly idealised models of the situation, or more realistic non-linear cases where unitary evolution will not in fact take place \cite{Ellis (2012), Ellis (2023)}. 

Moreover, any quantum description that neglects the black hole interior will always be non-unitary because it traces over states in the interior and thus uses a density matrix, not a wave function.  But to properly include the interior one must include the singularity.  If the singularity exists even in a quantum gravity description of black holes then information can be lost because singularities can destroy information.  If the singularity does not exist in quantum gravity then the answer to the information paradox is likely to be tied to the nature of whatever quantum gravity replaces the singularity with.  In any case that description is unlikely to be a single classical spacetime with an associated quantum state for the matter fields.

\subsection{Black hole evaporation}
When a black hole emits radiation through the Hawking effect those of us who stay outside the black hole will describe the radiation by tracing over the state in the interior of the black hole and thus obtain a density matrix.  But then once the black hole evaporates completely, it sounds like we're just left with a density matrix. That doesn't sound like something quantum mechanics should do, allowing a pure state to evolve into a density matrix.  So what (if anything) was missing from the original calculation of black hole evaporation? 

The original calculation was done in the semiclassical approximation in which gravity is treated classically (using general relativity) and everything else is treated using quantum mechanics.  So clearly one thing that was missing from the original calculation is the effects of quantum gravity.  The singularity deep in the interior of the black hole is where quantum gravity effects are important, but it is far from where the radiation is emitted, so instead string theorists decided to concentrate on the effects of quantum gravity near the event horizon.  

On the face of it, this sounds like a terrible idea (and in our opinion it really is a terrible idea), but that's the path most researchers decided to pursue and have pursued to this day. What's disturbing to us about regarding the horizon and its vicinity as a place of strong gravity is that it seems like a giant step backwards from the hard-won geometric insights of the global structure program.

Recall that the coordinate components of the Schwarzschild metric are singular in the usual Schwarzschild coordinate system.  But one of the insights of the global structure program is that geometry is important and coordinates aren't.  So the horizon is fine, but the Schwarzshild time coordinate $t$ is not: it goes to infinity on the horizon.  

More generally, the horizon is a null surface, but if you regard a null surface as a limit of a family of timelike surfaces, then in the limit the geometry of the surface becomes singular.  One can consider a ``stretched horizon''\cite{Susskind (1993)} a timelike surface very near the event horizon and thus a place where (if you don't think geometrically) it sounds like the effects of quantum gravity might be strong.

Ironically the stretched horizon is almost identical to Kip Thorne's black hole membrane paradigm \cite{Price and Thorne (1988)}.  But Thorne wasn't so much trying to generate new insight as trying to make black holes user friendly to astrophysicists who (Thorne thought) wouldn't have the time or inclination to absorb the insights of the global structure program.  The irony is that now that there is plenty of astrophysical data about black holes, astrophysicists have shown themselves nimble enough to understand black holes just fine.  So the main use of membranes (a.k.a stretched horizons) is to those who want to think of black hole horizons as extreme environments.  

Another difficulty with concentrating on the horizon is that the black hole interior is a four dimensional spacetime, while the horizon is a three dimensional null surface.  Are there really enough degrees of freedom in the horizon to account for all the stuff we trace over in the interior?  For this to be the case, nature would have to satisfy some sort of ``holographic principle''\cite{Susskind (1993)} by which a surface has as much information as the volume it surrounds.  

At the time the holographic principle was proposed, no system was known that satisfied it, but the proposed principle motivated the search for such a system and in due course one was found: the \href{https://en.wikipedia.org/wiki/AdS/CFT_correspondence}{AdS/CFT} correspondence \cite{Maldacena (1998)}.  Here AdS stands for string theory in a spacetime that is asymptotic to anti-de Sitter spacetime, and CFT stands for a supersymmetric conformal field theory on the boundary at infinity of the spacetime.  The correspondence is that the predictions of one theory in the limit of strong coupling are also those of the other theory in the limit of weak coupling.  

AdS/CFT is an interesting mathematical system.  But we have serious doubts as to whether it describes the physics of our universe.  To begin with, our spacetime is not asymptotically anti-de Sitter, which has a negative cosmological constant $\Lambda$ contra to astronomical observations,  and the standard model is not a supersymmetric conformal field theory.  Moreover, from the geometric point of view there is something disturbing about using anti-de Sitter spacetime as a model: recall that in addition to energy conditions, the global structure program also uses causality conditions, in particular the existence of a Cauchy surface.  Anti-de Sitter spacetime violates the energy condition by having a negative cosmological constant.  In addition it violates the causality conditon by not having a Cauchy surface.  This is because there are accelerated observers who can get to infinity in a finite AdS time.  Thus the boundary at infinity (which is needed for there to be any correspondence) is there only because from the geometric point of view anti-de Sitter spacetime is pathological. It is not the universe in which we live. 

The holographic principle also gave rise to an argument that information gets out of an evaporating black holes as follows: If an evaporating black hole is equivalent to a conformal field theory, then since 
 in field theories pure states evolve to pure states,  therefore the final state of an evaporating black hole is a pure state.  Since we're not convinced that nature obeys a holographic principle, we're not convinced by holographic arguments involving extra dimensions about black hole evaporation .

\subsection{The Firewall and ER = EPR}
Nonetheless, even if one is convinced by holographic arguments there remains the following question: exactly how does the information get out of the black hole?  Several mechanisms were considered, but finally one group \cite{Almheiri et al (2013)} (hereafter known as AMPS) produced a no-go result: any mechanism effective enough to get the information out would actually give rise to a huge amount of stress-energy on the horizon (a ``firewall'').  No mechanism would work.  

Though the AMPS paper is often thought of as predicting a firewall, its structure is actually that of a standard proof by contradiction:  one of their assumptions (in addition to information getting out) is the absence of huge stress-energy in the vicinity of the horizon (i.e. the absence of a firewall).  When their reasoning then leads to the presence of a firewall they have obtained a contradiction with one of their original assumptions thus showing that the original set of assumptions can't obtain. 

From our point of view, the result of the AMPS paper is both welcome and unsurprising.  Everything that we know about quantum mechanics and null surfaces leads us to believe that quantum information can't emerge from behind a null surface.  

One really nice feature of the AMPS paper is that it also gives the reason why information can't get out: the quantum no cloning property. Though for many purposes, we get to ignore black hole interiors, they contain a lot of quantum information.  For that information to also get out, there would have to be two copies of it: one inside the black hole and one outside, which would violate quantum no-cloning.

We mention in passing that though we are quite fond of no-go theorems, not everyone feels that way.  For example, Werner Israel liked to refer to people who prove no-go theorems as ``mathe-morticians'' by which he meant people who use their mathematical skills to bury an idea.  For those who dislike either no-go theorems in general, or just a particular no-go theorem, there is always a way to evade the consequences of the theorem.  This is because all theorems have assumptions, so if you don't like the conclusions of the theorem, you can evade the conclusions by considering a situation where one or more of the assumptions don't obtain.

 And now we (finally) come to the reason for ER=EPR: it is a way to evade the no-go result of AMPS.  Essentially the evasion works as follows: suppose that all black holes are really wormholes.  Then there are enough degrees of freedom in the other spacetime that we could imagine people in the other spacetime doing things that would cancel out the firewall that would occur if the black hole were just a black hole.  This still doesn't tell us anything about how the information is supposed to get out, but at least it's no longer impossible. 

 On the formal level, the assumption of AMPS that is being dropped is that we are dealing with the evaporation of a black hole (rather than a wormhole).  While formally this sort of maneuver is perfectly acceptable, from the geometric point of view (and it seems to us also the physical point of view) the statement ``maybe there just aren't any black holes'' seems like a pretty extreme maneuver. It will certainly surprise all those who study black holes from an astrophysical viewpoint \cite{Begelman and Rees (2020)}

\section{Geometric and non-geometric approaches; ``computational complexity''}

Though we find the question of black hole evaporation very interesting, we are dismayed by how much of it is done in an explicitly anti-geometric way.  We note that it is a little over 50 years since  
Penrose and Hawking and Geroch and the other researchers of the global structure program brought new geometric insight to the relativity community \cite{Penrose (1965),Hawking and Ellis (2023),Senovilla and Garfinkle (2015)} and the
physics community in general.  We are concerned that much of that insight will be lost: drowned in a deluge of anti-geometric attitudes.

We close with two last examples. The first has to do with the concept of spacetime and foliations.  Physics is about dynamics: the development of a system as a function of time.  But to treat general relativity this way, one needs to divide spacetime up into space and time: that is one needs a foliation.  One important insight of the global structure program is that the spacetime is the physics: the choice of foliation is completely arbitrary to be chosen by the user for their own convenience.  This insight was crucial for understanding black holes since otherwise one might be tempted to look at the Schwarzschild spacetime only in the foliation by constant Schwarzschild coordinate time slices.  One would then be misled into thinking that it takes forever to fall into a black hole since the Schwarzschild time slices never reach the future event horizon.  
In contrast to this geometric insight consider the work of \cite{Susskind (2014)} which treats black hole interiors and speculates about what their field theory duals might be. 
This work makes a particular choice of foliation: maximal slicing.  Much of the paper is taken up with speculation on a supposed dual relation between growth of volume of the slices and ``growth of complexity'' on the field theory side.  This growth of volume of maximal slices inside black holes is well known in numerical relativity where it is associated with the ``collapse of the lapse'' i.e. the fact that the slices slow down inside the black hole and stay well away from the singularity.  But maximal slicing is just one of many slicing conditions that could be chosen.  So to single out this particular slicing as the right slicing, the one whose dynamics has physical meaning, is to adopt a specifically anti-geometric attitude. 

 The second example has to do with a growing body of work on the idea of  ``computational complexity'' and black holes \cite{Susskind (2016),Aaronson (2016),Haferkamp et al (2022),Chapman and Policastro  (2022)} which derives from the literature on firewalls plus that on the AdS/CFT formalism as well as that on ER bridges. \cite{Susskind (2016)}   states (\S3),\begin{quote}
     ``\textit{It is commonly believed that black holes can be
modeled as systems of N qubits with N is of order the entropy of the black hole}.''
 \end{quote} We are completely baffled by such statements. 
 The concept of a physical qubit\footnote{\hyperref{https://en.wikipedia.org/wiki/Qubit}{qbits} : ``A qubit is a two-state (or two-level) quantum-mechanical system, one of the simplest quantum systems displaying the peculiarity of quantum mechanics'' 
 (Wikipedia).} only arises in the context of a quantum computer, where they are controlled by algorithms 
 as explained in \cite{DiVincenzo (2000)}:
\begin{quote}
`\textit{`A quantum algorithm is typically specified as a sequence of unitary transformations U1, U2, U3,..., each acting on a small number of qubits...  The most straightforward transcription
of this into a physical specification is to identify Hamiltonians which generate these unitary transformations''.} 
\end{quote}
In the context of a black hole, precisely what are the qubits, and what generates these unitary transformations? Where does the relevant algorithm come from? 
Why is any such concept needed or relevant? 
As a consequence of the black hole uniqueness theorems 
\cite{Robinson (2009)}, the geometry of a black hole ${\cal B}$ is classically described by just  three parameters: its mass $m$, rotation $a$, and charge $e$, so its number of degrees of freedom ${\cal D(B)}$ is 3. That is a measure of its complexity.  If we consider quantum gravity, the graviton is presumably a massless spin-2 boson (see \cite{Weinberg (1972)} and references therein). There is nothing remotely resembling qubits in sight. Solid physics has given way to what appears to be pure fantasy: layer upon layer of unsubstantiated concepts being invoked\footnote{See \hyperref{https://scottaaronson.blog/?p=6599}{``On black holes, holography, the Quantum Extended Church-Turing Thesis, fully homomorphic encryption, and brain uploading''} by Scott Aaronson to get the flavour.}.

The actual geometry of black holes (\cite{Hawking and Ellis (2023)}:\S5.5, see Fig. 1) has nothing to do with AdS/CFT, which is the centre of such speculation, or qbits. It is perfectly well described by General Relativity as usually understood. As for holography and general relativity proper, there are cases where the idea can be understood in a sensible way that does not involve AdS/CFT. It is basically a restatement of the null initial value problem \cite{Goswami and Ellis (2017)} but runs into problems in the cosmological case  because of cusps and folds in null surfaces \cite{Tavakol and Ellis (1999)}.\\

\section*{Acknowledgement}

DG was supported by NSF Grant PHY-2102914 and GE by the UCT Science Faculty Research Fund.

\vspace{0.1in}
    Version 2023/11/12
\end{document}